\documentclass{mem}

\usepackage{natbib}
\usepackage{txfonts}
\usepackage{balance}
\usepackage{graphicx}
\usepackage[a4paper]{hyperref}
\idline{00}{81}

\newcommand{\ksk}{km~s$^{-1}$~kpc$^{-1}$}

\begin{document}

\title{Ordered and chaotic spirals in disk galaxies}

\subtitle{}

\author{P. A. Patsis\inst{1,2} 
  \and C. Kalapotharakos\inst{1} }

\offprints{Panos A. Patsis; \email{patsis@academyofathens.gr}}

\institute{Research Center for Astronomy,
  Academy of Athens, Athens, Greece
\and
  European Southern Observatory, Garching bei M\"{u}nchen, Germany
}

\authorrunning{Patsis and Kalapotharakos}

\titlerunning{Chaotic spirals}

\abstract{The pattern speeds of  spiral galaxies 
are closely related to the flow of material in their disks.  Flows
that follow the `precessing ellipses' paradigm (see e.g., Kalnajs
1973) are likely associated with slowly rotating spirals, which have
corotation beyond their end. Such a flow can be secured by material
trapped around stable, elliptical, x$_1$ periodic orbits precessing as
their Jacobi constant varies. Contrarily, if part of the spiral arms
is located at a corotation region then the spiral structure has to
`survive' in chaotic regions. Barred-spiral systems with a single
pattern speed and a bar ending before, but close to, corotation are
candidates for having spirals supported by stars in chaotic motion. In
this work we review the flows we have found in response models for
various types of spiral potentials and indicate the cases, where order
or chaos shapes the observed morphologies.

}

\maketitle{}

\section{Introduction}

In the present paper we call \textit{ordered} spirals those that have
as building block a set of stable periodic orbits. The standard
example are the spirals in Contopoulos \& Grosb{\o}l (1986) models,
which end at their inner 4:1 resonance. In this type of models the
spiral structure terminates at a distance beyond which there are no
more quasiperiodic orbits trapped around stable, elliptical periodic
orbits of the x$_1$ family. On the other hand we will call
\textit{chaotic} spirals those that we believe are constituted from
stars in chaotic motion. Kaufmann \& Contopoulos (1996) argued for the
first time that part of the spirals in barred-spiral systems with a
single pattern speed are chaotic. The galactic morphologies we study
are grand design independently of whether their spirals are ordered or
chaotic. We use response models in trying to model these
morphologies. The potentials we use are either analytic forms that
describe the structures we study, or potentials that have been
directly determined from near-infrared observations of specific
galaxies.

\begin{figure*}[t!]
\begin{center}
\resizebox{\hsize}{!}{\includegraphics[scale=0.55]{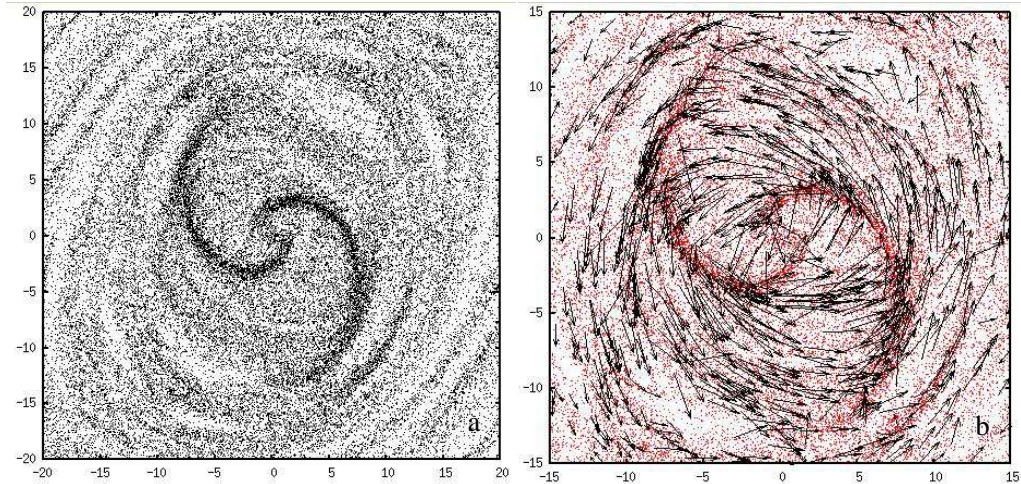}}
\end{center}
\caption{\footnotesize
  The flow of stars in a time-independent stellar response model of
  Contopoulos \& Grosb{\o}l (1986) type. (a) The response
  morphology. Corotation is at $\approx$ 24~kpc and at the end of the
  strong spirals, at $\approx$ 12.5~kpc, is the inner 4:1
  resonance. (b) The velocity field of the model indicated by arrows.}
\label{normal}
\end{figure*}

\section{Normal spiral galaxies}

Normal spiral galaxies have no or very weak bar components. Orbital
models (Patsis et al. 1991) and gaseous response models (Patsis et
al. 1994, 1997a) have shown that open, grand design, \textit{normal}
spiral galaxies are best modeled by slowly rotating patterns. Then
corotation is beyond the end of the spiral arms, or at least beyond
the end of a characteristically bisymmetric part of the arms. Since
the galaxies we refer to are grand design, their symmetric part
dominates their morphology. Nevertheless, especially in gaseous
response models, bifurcations of the arms appear at the end of the
symmetric part. Also off-phase extensions of the spirals with respect
to the imposed potential minima may appear beyond the point at which
symmetry breaks. The inclusion of an $m=1$ component in phase with the
main $m=2$ spiral and with the same pattern speed, is crucial for
modeling the observed structures (see Fig.~4 in Patsis et
al. 1997a). The point at which the symmetric part of the spirals ends
can be used for the estimation of their pattern speed, since it is
associated with the inner 4:1 resonance (Contopoulos \& Grosb{\o}l
1986). Very frequently we observe there a characteristic bifurcation
of the arms.

In the last 12 years, stellar and gaseous response models for normal
spirals, or for spirals that have their own pattern speeds in
barred-spiral systems, consistently put corotation beyond the end of
the symmetric part of the spirals and in some cases even beyond the
end of the spiral structure altogether (Mulder \& Combes 1996; Kranz
et al. 2001; Bissanz et al. 2003; Kranz et al. 2003; Pichardo et
al. 2003; Martos et al. 2004; Martos \& Yanez 2005; Vorobyov 2006;
Vorobyov \& Shchekinov 2006; Minchev \& Quillen 2008; Patsis et
al. 2009). All these models strongly indicate that the flow in normal
spirals follows `precessing ellipses'. Since the arms extend inside
corotation, they are located in a disk region, where order dominates
and thus structures can be built by quasi-periodic orbits trapped
around stable periodic orbits.

Fig.~\ref{normal} shows a typical stellar response model to an imposed
bisymmetric potential of the type used by Contopoulos \& Grosb{\o}l
(1986). Only an $m=2$ term is included. Corotation is at 24~kpc. The
main spirals end at their inner 4:1 resonance, inside corotation
(Fig.~\ref{normal}a) and the flow of the stars is along `precessing
ellipses' (Fig.~\ref{normal}b).

\begin{figure*}[t!]
\begin{center}
\resizebox{\hsize}{!}{\includegraphics[scale=0.7]{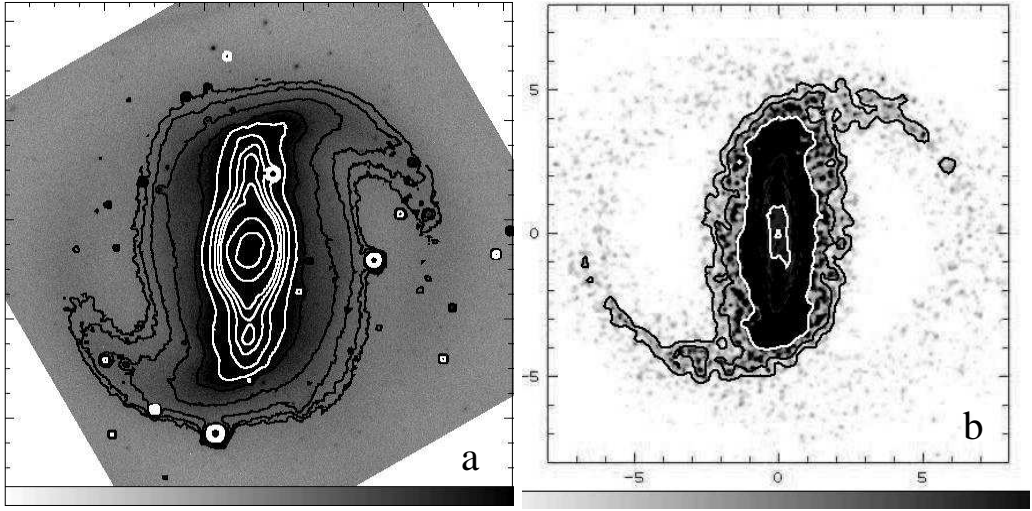}} 
\end{center}
\caption{\footnotesize
  (a) An $R$-band image of NGC~4314 showing the boxiness of the outer
  isophotes of the bar. We observe the limited azimuthal extent of the
  spirals. (b) A stellar response model of the Quillen et al. (1994)
  potential. The pattern rotates with $\Omega_{\rm p}=38.23$~\ksk. The
  model reproduces the basic morphological features of the galaxy.}
\label{n4314}
\end{figure*}

\section{Barred spiral galaxies}

Bars end close, but before, corotation (Contopoulos 1980). In
barred-spiral systems with one pattern speed and the observed spiral
arms attached to the ends of the bar, the spirals have to cross the
corotation region where chaos dominates. We present the orbits that we
find to support such a spiral structure in two types of barred-spiral
potentials. Both of them result from near-infrared observations of real
galaxies. The values of the pattern speeds ($\Omega_{\rm p}$) we use
in the examples we present below does not necessarily give the best
fit of the galaxy morphology in all cases. We just use template models
to present the orbital behavior in barred-spiral systems with boxy
bars and with bars with ansae morphology.

\subsection{Boxy bars} 

\begin{figure*}[t!]
\begin{center}
\resizebox{\hsize}{!}{\includegraphics[scale=0.4]{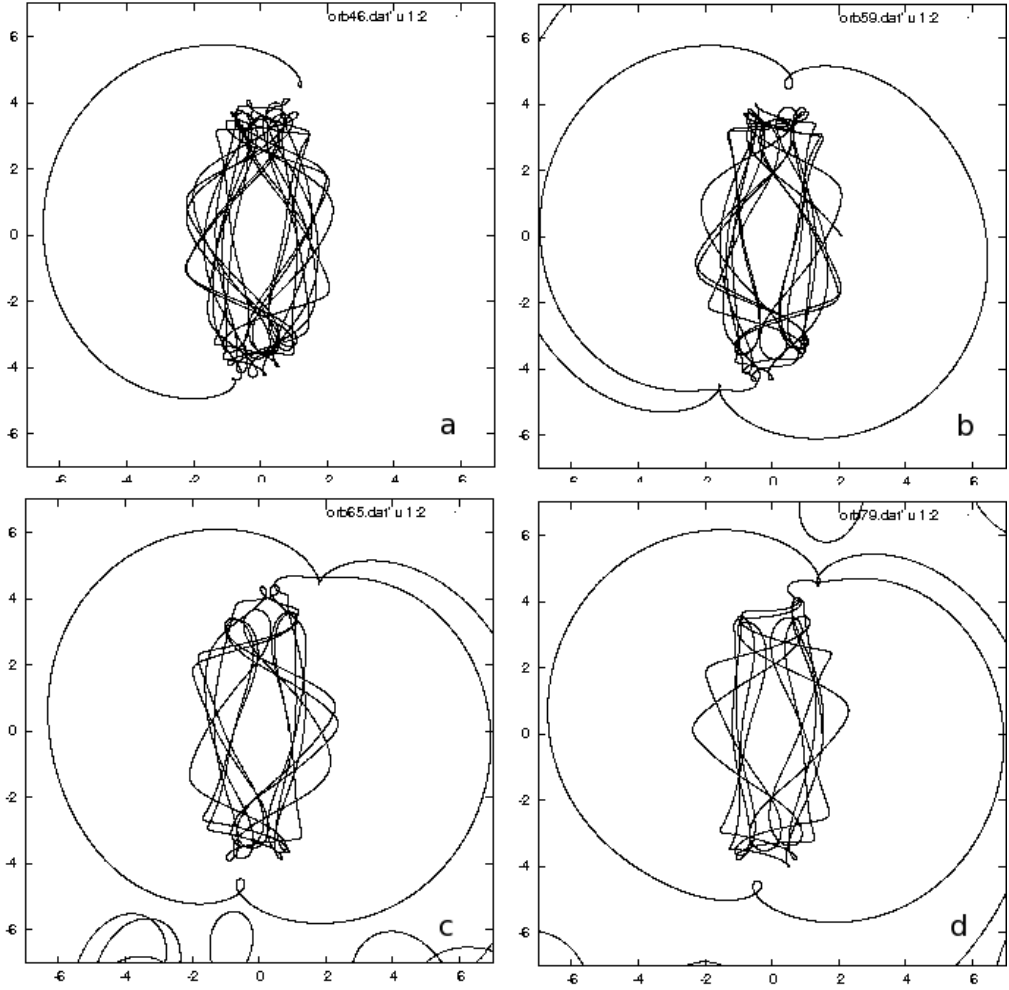}}
\end{center}
\caption{\footnotesize
  Chaotic orbits of particles that support the spiral structure we
  observe in Fig.~\ref{n4314}. During the time they spend in the bar
  region they have a rectangular-like morphology, typical of orbits at
  the 4:1 region of bars. They are integrated for 10 bar periods.}
\label{orbits}
\end{figure*}

We call `boxy' the bars that have, on the plane of the galaxy, outer
isophotes with rectangular-like shape. A standard example is the
nearly face-on, early type barred-spiral galaxy NGC~4314. In
Fig.~\ref{n4314}a we give an R-image (Gadotti \& de Souza 2006) with
overplotted isophotes showing the boxiness in discussion. The spiral
structure appears as a continuation of the bar and is confined
azimuthally in an angle less than $\pi/2$. Patsis et al. (1997b), have
shown that the stellar orbits associated with the observed boxy
isophotes are chaotic. The potential used for this study has been
estimated by Quillen et al. (1994) from the $K$-image of
NGC~4314. Stellar responses are similar for a range of $\Omega_{\rm
p}$ values $38<\Omega_{\rm p}<45$~\ksk. The pattern in the response
model of Fig.~\ref{n4314}b rotates with $\Omega_{\rm
p}=38.23$~\ksk. Patsis (2006) has shown that the chaotic orbits that
reinforce the outer boxy isophotes of the bar are practically the
same, which sustain also the spirals that emerge from the ends of the
bar. They belong to a population that visits both the bar and the disk
area further out. During the time these chaotic orbits visit the bar
region, they have a morphological resemblance with quasi-periodic
orbits trapped around rectangular-like 4:1 type stable periodic orbits
we find in generic barred potentials (Contopoulos 1988). However, in
the potential we study, this family is stable over a tiny energy
interval, and the space their invariant curves occupy in the phase
space is tiny (Patsis et al. 1997b). In Fig.~\ref{orbits} we give
characteristic examples of orbits of particles that are located on the
spiral arms in Fig.~\ref{n4314}b.

\subsection{Bars with ansae}

\begin{figure}[t!]
\begin{center}
\resizebox{\hsize}{!}{\includegraphics[scale=0.7]{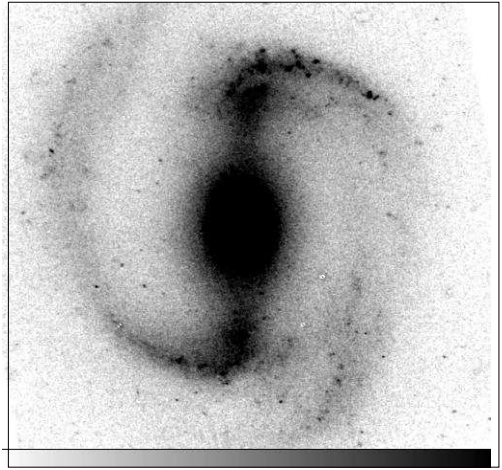}} 
\end{center}
\caption{\footnotesize
  The deprojected $K$-band image of the barred-spiral galaxy NGC~1300,
  which we used for our calculations in potentials with ansae type
  bars.}
\label{n1300bar}
\end{figure}

Another type of barred-spiral morphology we have studied has an ansae
type bar. The potential we used in this case is from an estimation of
the potential of NGC~1300, from a $K$-band image of this galaxy
(Kalapotharakos et al. 2008). For the figures we present below we have
used as inclination and position angles for the bar $42\fdg2$ and
$100\fdg6$ respectively. The $K$-band image of the galaxy deprojected
with these values is given in Fig.~\ref{n1300bar}.

\begin{figure*}[t!]
\begin{center}
\resizebox{\hsize}{!}{\includegraphics[scale=0.45]{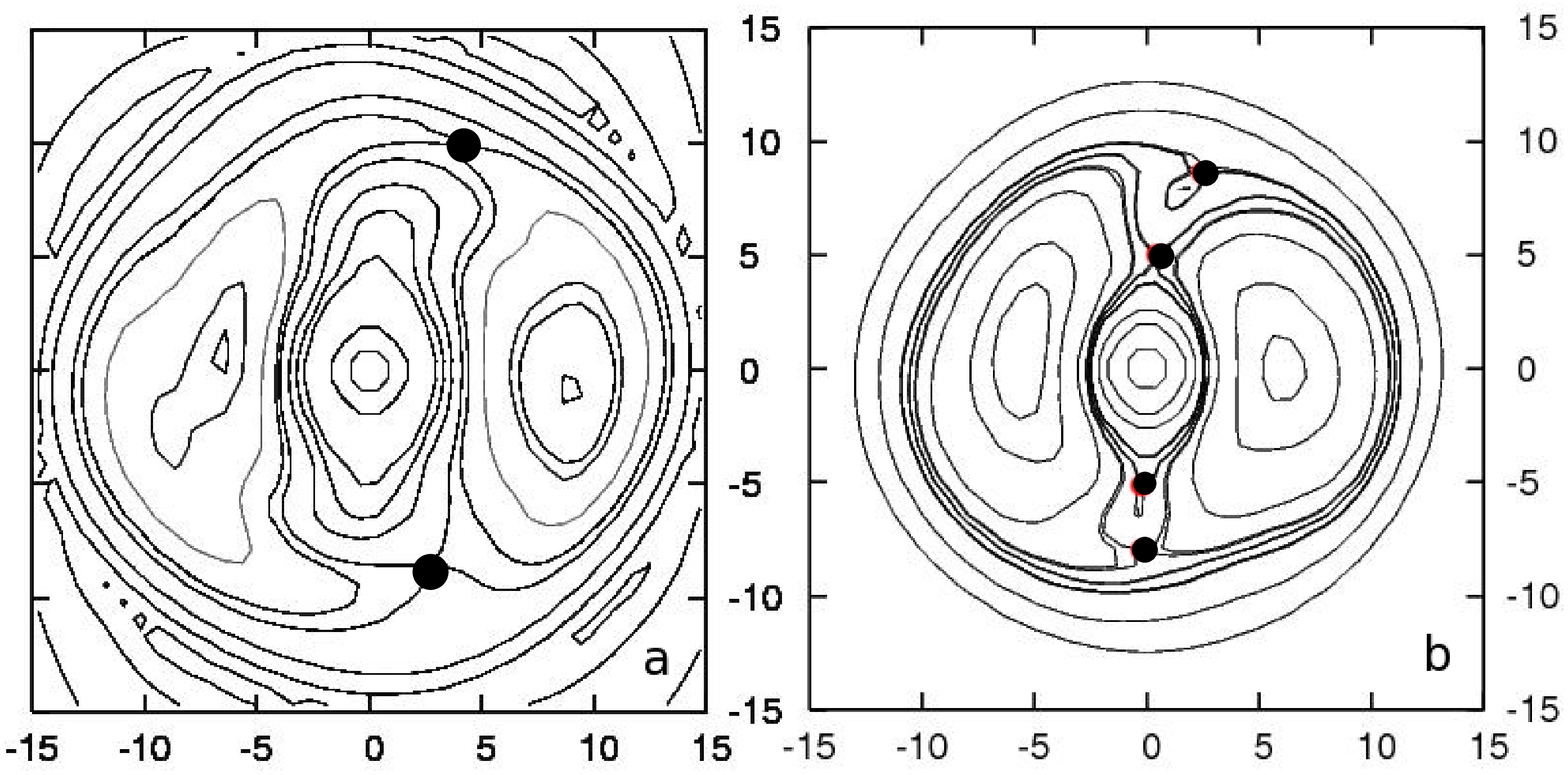}}
\end{center}
\vspace*{-0.5cm}
\caption{\footnotesize
  Equipotentials of effective potentials in models for NGC~1300. In
  (a) $\Omega_{p}$=21~\ksk\ and in (b) $\Omega_{p}$=26~\ksk. Heavy dots
  indicate the unstable Lagrangian points in the two models.}
\label{effpot}
\end{figure*}

It is interesting that in these models we find qualitatively different
effective potentials for nearby values of $\Omega_{\rm p}$. In
Fig.~\ref{effpot} we give the isocontours of two effective potentials
we studied. In Fig.~\ref{effpot}a, $\Omega_{\rm p}$=21~\ksk, while in
(b) $\Omega_{\rm p}$=26~\ksk. In the case with $\Omega_{\rm
p}$=26~\ksk\ the model has two sets of unstable Lagrangian points, and
the ansae morphology appears also in the isocontours. Nevertheless,
when $\Omega_{\rm p}$=21~\ksk, we have a more conventional
configuration with two unstable Lagrangian points close to the ends of
the bar. The models are asymmetric, because in the Fourier analysis we
have taken into account also the odd terms (Kalapotharakos et
al. 2008). Despite the differences these two models give similar
stellar responses. Again here we consider the orbits that support the
spirals found. They are chaotic and their typical morphologies can be
seen in Fig.~\ref{13orbits}. As in the previous barred-spiral case we
have orbits that visit both the bar region as well as the disk beyond
its end. During the time they stay at the bar region they develop a
morphology which again has a 4:1 resonance character. However, this
morphology is not rectangular-like, but like another type of 4:1
orbits. These are elliptical-like with loops at their apocentra
(Contopoulos 1980; Patsis et al. 1997b). These chaotic orbits shape
also the ansae, performing a number of loops at the corresponding
regions.

\section{Other cases}

\subsection{Barred galaxies with extended spiral structure}

In both barred-spiral galaxies we have studied above, the spiral arms
are radially and/or azimuthally confined to short
distances. Especially in NGC~1300, the spiral structure is very
asymmetric. The right arm, as depicted in Fig.~\ref{n1300bar},
practically breaks in two parts. In NGC~4314, the extent of the
spirals azimuthally does not exceed $\pi/2$. This is a morphology that
differs from the logarithmic spirals of nearly normal grand design
spiral galaxies, where we do not observe major gaps (Grosb{\o}l \&
Patsis 1998). A case of a barred-spiral system with a radially
extended spiral structure without major gaps and discontinuities along
the arms, has been studied by Boonyasait et al. (2005) and Patsis et
al. (2009) using a potential estimated from near-infrared observations
of NGC~3359. For these `well organized' spiral arms the best matching
of the flow at their region has been obtained by a
`precessing-ellipses' flow. The spiral structure in the barred-spiral
galaxy NGC~3359 is supported by regular orbits. This is an indication
that there is a correspondence between morphological features of the
arms (extent, presence of gaps etc.) and the dynamical mechanism that
sustains them in each individual case.

\subsection{Normal double spirals}

As we have noted in Sect.~2, in both gaseous and stellar response
models of normal open spirals we find the density response maxima
along the imposed potential minima only up to the inner 4:1
resonance. Beyond that distance we find in some cases off-phase
extensions between the 4:1 resonance and corotation in gaseous
models. Nevertheless, if we fine tune our models, so that we have a
perturbing force of the order of 10\% of the axisymmetric background
at corotation, and populate the region beyond corotation with enough
particles, we observe another set of spirals \textit{beyond}
corotation. These spirals are weak and do not have the pitch angle of
the imposed potential. This is the case depicted in Fig.~\ref{dble}.

The strong spirals we observe in Fig.~\ref{dble}a end at the 4:1
resonance and are supported by regular orbits. The weak spirals beyond
corotation, emerge close to the unstable Lagrangian points of the
system and follow a chaotic flow, which is shown in Fig.~\ref{dble}b.
Double spirals in response models with a gap at the corotation region
are obtained also by Vorobyov (2006). These morphologies are not
frequently observed. However, grand design spiral galaxies like
NGC~1566 and NGC~5248 show this morphology and are candidates of
hosting both ordered and chaotic spirals.

\section{Conclusions and discussion}

The orbital response models that we have studied lead us to the
following conclusions:

\begin{figure*}[t!]
\begin{center}
\includegraphics[scale=0.5]{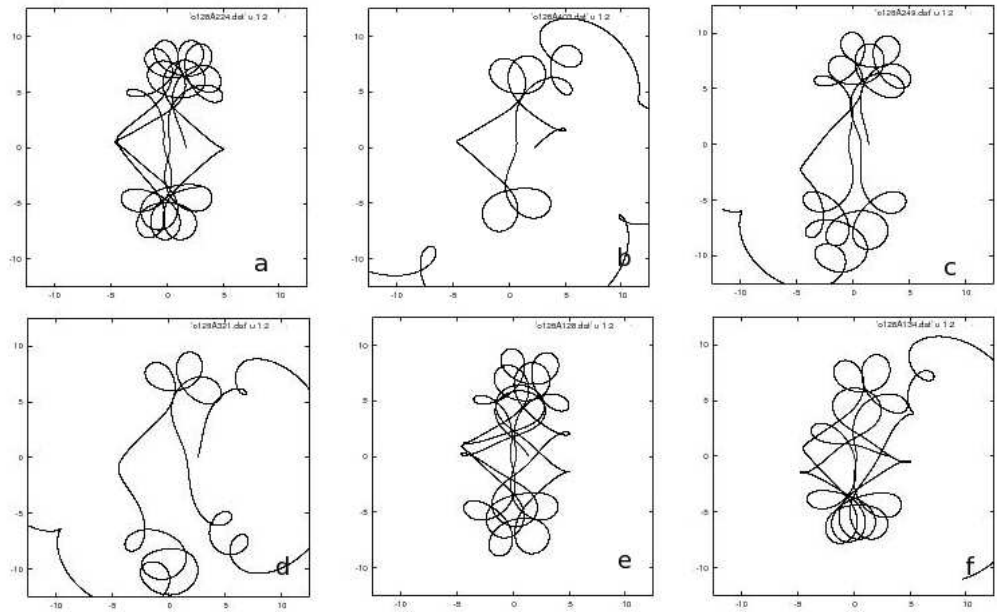}
\end{center}
\vspace*{-2.3cm}
\caption{\footnotesize
  Typical orbits that support barred-spiral morphologies with
  ansae. They support both the spirals as well as the ansae
  features. They are integrated for 10 bar periods.}
\label{13orbits}
\end{figure*}

\begin{figure*}[t!]
\begin{center}
\resizebox{\hsize}{!}{\includegraphics[clip=true]{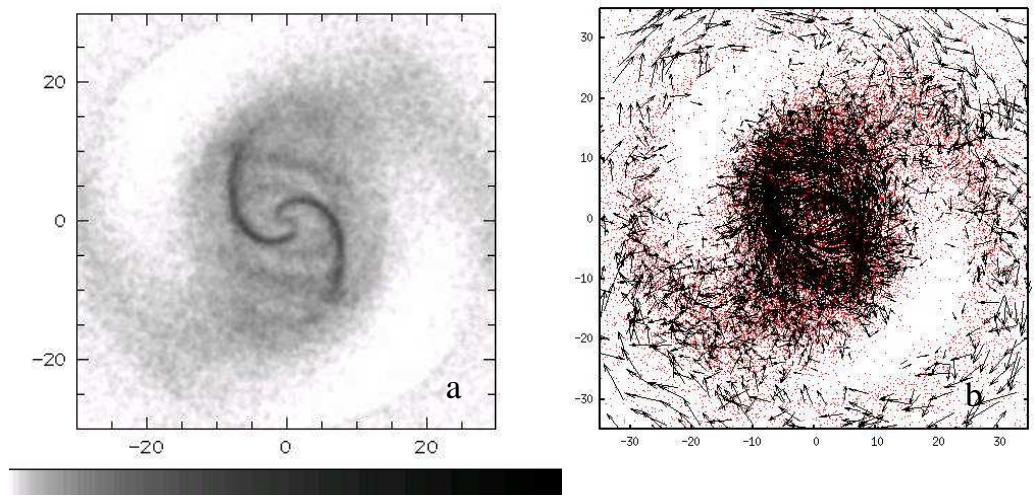}}
\end{center}
\caption{\footnotesize
  A model of a double normal spiral. In (a) we observe the inner
  strong and the outer weak spiral. In (b) we give the velocity field
  with arrows, so that we can see the flow along the spirals beyond
  corotation.}
\label{dble}
\end{figure*}

\begin{enumerate}

\item Response models of (nearly) normal 
  grand design spiral galaxies indicate that the spiral arms in this
  type of galaxies rotate slowly. Corotation is beyond the end of the
  bisymmetric part of the spirals and the grand design is supported by
  regular orbits. There are several published simulations in the
  literature that reproduce this result. As soon as the perturbing
  forces are larger than 5\% of those of the axisymmetric background,
  the nonlinear phenomena described by Contopoulos \& Grosb{\o}l
  (1986) are present in the models. The final response morphology is
  obtained after a few pattern rotations. This means that these models
  can predict galactic morphologies that last for a few Gyr.

\item Barred-spiral systems with a single pattern speed 
  may have spirals that are supported by chaotic orbits. In the cases
  we have studied and we have found that the spirals are supported by
  particles in chaotic motion (NGC~4314, NGC~1300), the spiral arms
  are either confined azimuthally in their extent, or asymmetric with
  gaps. Contrarily in a barred-spiral galaxy with a well described
  spiral structure (NGC~3359) our results are in agreement with a flow
  of regular orbits. The conditions under which barred-spiral systems
  have ordered or chaotic spirals needs further investigation. We
  mention the work of other groups based on $N$-body models and
  orbital theory that present models of barred-spiral systems with a
  well developed spiral structure consisting entirely by particles in
  chaotic motion (see Efthymiopoulos et al. 2008 and references
  therein, as well as Romero-Gomez et al. 2008 and references
  therein).

\item The chaotic orbits that support the spirals 
  belong to a population that visits both the bar and the disk region
  beyond corotation. In the two examples we studied we find that they
  are the same orbits that shape the outer structure of the bars. They
  support the boxy and the ansae type morphologies of our models. This
  means that they belong to a very important orbital population for
  understanding the shapes of the bars. On a surface of section we
  find the initial conditions in a chaotic sea, that support the bar
  if integrated for a certain number of bar revolutions (of the order
  of 10). During the time these orbits stay at the bar region, they
  follow trajectories with a morphology similar to those of 4:1
  resonance regular orbits trapped around stable periodic orbits. This
  gives to the bars shapes resembling periodic orbits at the 4:1
  resonance region.
\end{enumerate}

\begin{acknowledgements}
PAP thanks ESO for an invitation to visit ESO Garching for two months
during the summer of 2008, where this work has been completed. We
thank Prof. G. Contopoulos for valuable comments.
\end{acknowledgements}

\bibliographystyle{aa}

\begin{thebibliography}{}

\bibitem[{Bissantz et al.(2003)}]{beg03}
  Bissantz, N., Englmaier, P., \& Gerhard, O. E. 2003, MNRAS, 340, 949

\bibitem[{Boonyasait et al.(2005)}]{bpg05}
  Boonyasait, V., Patsis, P. A., \& Gottesman, S. T. 2005, Ann. New
  York Acc. Science, 1045, 203

\bibitem[{Contopoulos(1980)}]{c80}
  Contopoulos, G. 1980, A\&A 81, 198

\bibitem[{Contopoulos(1988)}]{c88}
  Contopoulos, G. 1988, A\&A 201, 44

\bibitem[{Contopoulos \& Grosb{\o}l(1986)}]{cg86}
  Contopoulos, G., \& Grosb{\o}l, P. 1986, A\&A, 155, 11

\bibitem[{Efthymiopoulos et al.(2008)}]{etkg}
  Efthymiopoulos, C., Tsoutsis, P., Kalapotharakos, C., \& Contopoulos
  G. 2008, in Chaos in Astronomy, ed. G. Contopoulos, \& P. A. Patsis
  (Springer, Berlin), 173

\bibitem[{Gadotti \& de Souza(2006)}]{gs06}
  Gadotti, D. A., \& de Souza, R. E. 2006, ApJS 163, 270

\bibitem[{Grosb{\o}l \& Patsis(1998)}]{gp98}
  Grosb{\o}l, P., \& Patsis, P. A. 1998, A\&A 336, 840

\bibitem[{Kalapotharakos et al.(2008)}]{kpg08}
  Kalapotharakos, C., Patsis, P. A., \& Grosb{\o}l, P. 2008, submitted

\bibitem[{Kalnajs(1973)}]{k73}
  Kalnajs, A. 1973, PASA 2, 174

\bibitem[Kaufmann \& Contopoulos(1996)]{kc96}
  Kaufmann, D. E., \& Contopoulos, G. 1996, A\&A, 309, 381

\bibitem[{Kranz et al.(2001)}]{ksr01}
  Kranz, T., Slyz, A., \& Rix, H.-W. 2001, ApJ, 562, 164

\bibitem[{Kranz et al.(2003)}]{ksr03}
  Kranz, T., Slyz, A., \& Rix, H.-W. 2003, ApJ, 586, 143

\bibitem[{Martos et al.(2004)}]{mhymp04}
  Martos, M., Hernandez, X., Yanez, M., \& Pichardo, B. 2004, MNRAS,
  350, L47

\bibitem[{Martos \& Yanez(2005)}]{my05}
  Martos, M., \& Yanez, M. 2005, in Magnetic Fields in the Universe:
  From Laboratory and Stars to Primordial Structures, AIP Conf. 784,
  ed. E. M. de Gouveia Dal Pino, G. Lugones, \& A. Lazarian (AIPC,
  New York), 362

\bibitem[{Minchev \& Quillen(2008)}]{mq08}
  Minchev, I., \& Quillen, A. 2008, MNRAS

\bibitem[{Mulder \& Combes(1996)}]{mc96}
  Mulder, P. S., \& Combes, F. 1996, A\&A 313, 723

\bibitem[{Patsis et al.(1991)}]{pcg91}
  Patsis, P. A., Contopoulos, G., \& Grosb{\o}l, P.  1991, A\&A, 243,
  372

\bibitem[{Patsis et al.(1994)}]{phcg94}
  Patsis, P. A., Hiotelis, N., Contopoulos, G., \& Grosb{\o}l, P.
  1994, A\&A, 286, 46

\bibitem[{Patsis et al.(1997a)}]{pgh97}
  Patsis, P. A., Grosb{\o}l, P., \& Hiotelis, N. 1997a, A\&A, 323, 762

\bibitem[{Patsis et al.(1997b)}]{paq97}
  Patsis, P. A., Athanassoula, E., \& Quillen, A. 1997b, ApJ 483, 731

\bibitem[{Patsis et al.(2008)}]{pkgb08}
  Patsis, P. A., Kaufmann, D. E., Gottesman, S., \& Boonyasait,
  V. 2009, MNRAS, 394, 142

\bibitem[{Pichardo et al.(2003)}]{pmme03}
  Pichardo, B., Martos, M., Moreno, E., \& Espresate, J. 2003, ApJ
  582, 230

\bibitem[{Quillen et al.(1994)}]{qfg94}
  Quillen, A. C., Frogel, J. A., \& Gonzalez, R. A. 1994, ApJ, 437,
  162

\bibitem[{Romero-G\'omez et al.(2008)}]{ramg08}
  Romero-G\'omez, M., Athanassoula, E., Masdemont, J. J., \&
  Garc\`{\i}a-G\'omez, C. 2008 in Chaos in Astronomy,
  ed. G. Contopoulos, \& P. A. Patsis (Springer, Berlin), 85

\bibitem[{Vorobyov(2006)}]{v06}
  Vorobyov, E. 2006, MNRAS, 270, 1046

\bibitem[{Vorobyov \& Shchekinov(2006)}]{vs06}
  Vorobyov, E., \& Shchekinov, Y. A. 2006, New Astronomy, 11, 240

\end{thebibliography}

\end{document}